\documentclass[twoside,12pt]{article}
\usepackage{epsfig}
\usepackage{amsbsy}
\usepackage{amssymb}
\usepackage{amsmath}

\def\Journal#1#2#3#4{{#1} {#2} (#4) #3 }

\def\PRD{{\em Phys. Rev.} D}

\def\APJ{{\em Ap. J.}}
\def\CQG{{\em Class. Quantum Grav.}}
\def\NAT{{\em Nature}}
\def\MNRAS{{\em Mon. Not. R. Ast. Soc.}}
\def\RPP{{\em Rep. Prog. Phys.}}

\newcommand{\be}{\begin{equation}}
\newcommand{\ee}{\end{equation}}
\newcommand{\bea}{\begin{eqnarray}}
\newcommand{\eea}{\end{eqnarray}}

\begin{document}

\title{ \vspace{1cm} The road to gravitational-wave astronomy}
\author{N.\ Andersson\\
\\
School of Mathematics, University of Southampton, Southampton, UK}
\maketitle
\begin{abstract}
Gravitational-wave astronomy is an area of great promise, yet to be realized.
While we are waiting for the first (undisputed!) direct detection of these elusive waves
it is useful to take stock and consider the challenges that need to be met if we want
this field to reach its full potential. This write-up provides a brief introduction to some of the key ideas
and the
current state-of-play, and lists
a range of modelling questions that need to be considered in the future.
\end{abstract}
%\eject
%\tableofcontents

\section{Trying to catch the wave}

Einstein's general theory of relativity predicts the existence of
gravitational waves; ripples in space-time  travelling at the speed of light. Attempts to detect these waves have so far not been
successful, but this should be about to change. After decades of development, the first generation of kilometre-scale
gravitational-wave interferometers have reached their designed sensitivity in a
broad frequency window. More than one year's worth of quality data was
taken during the fifth science run (S5) of the LIGO detectors \cite{Lrev} (together with the smaller British-German
GEO600 detector), and with the addition of the
French-Italian Virgo instrument located outside Pisa, we now have a network
of detectors searching for signals from astrophysical sources \cite{jim}.

It is not really surprising that, despite the advances in technology, gravitational waves
have not yet been detected. At the present level, the detectors are sensitive enough to
see unique events in the neighbourhood of our own Galaxy. Unfortunately, such
events are rare. Take supernovae, which occur only a few times per century in a
typical galaxy as an example. To secure detection, we ``simply'' need to reach further
out into the Universe. Exactly how far, we don't know at this point.
It is easy to work out the energy that has to be released in order for a given source to be detectable,
but very hard to provide a reliable model of the complex physics associated with most
gravitational-wave scenarios. What is clear is that we are dealing with very weak signals.
This is in obvious contrast with mainstream astronomy, where observations traditionally are made with a large signal-to-noise ratio.

One of the most recent papers of the joint LIGO-GEO-Virgo collaboration considers ``unmodelled'' burst sources \cite{burst}.
The analysis sets a limit on the energy required for a burst event to be detectable. Basically, the results demonstrate
that the detectors were sensitive enough to detect a Milky Way supernova, should one
have occurred during the observation run. These results add to a collection of upper limits, constraining the emission
from a range of sources from inspiralling compact binaries \cite{bin1} to (among others)
rotating deformed neutron stars \cite{def1} and from a stochastic gravitational-wave background
originating in the early Universe \cite{stoch}.

The present upper limits hardly challenge our understanding of the Universe.
From the general astrophysics point-of-view, the fact that the Crab pulsar releases less than a few
per cent of its energy as gravitational waves \cite{def1} is only mildly interesting. From the
perspective of gravitational-wave physics it is a milestone result. Constraining the level of gravitational-wave emission
to a fraction of that set by the observed pulsar spin-down, the Crab result hints at the
promise of future observations.

Gravitational-wave physics faces a number of challenges, ranging from the development
of detector technology to data handling techniques and theory modelling. To be
successful, we need to improve on all these aspects. At present, we have an impressive engineering
project, but in the future we want to turn this into ``astronomy''. The prospects for this are
good and we should not lose confidence at this point.
By 2015 the detector technology will have been upgraded, and the sensitivity
improved by an order of magnitude. The second generation of instruments will probe
a volume of space that is 1,000 times larger than is possible at present. According to
conservative population synthesis models, we should then be able to detect several
inspiralling binaries per year. Given that the ``bread and butter'' binary signals are well
understood and the data analysis algorithms are already (more or less) developed, we have good
reasons to expect this breakthrough.

The first direct detection of gravitational waves will
be tremendously exciting. It is also likely to motivate detailed consideration of a
wider range of astrophysical sources. This will put more emphasis on understanding the
physics involved and high-quality modelling of relevant astrophysical scenarios. Inevitably,
this will require an exchange of expertise with mainstream astronomers.
So far, the emphasis has very much been on detector
development and data analysis strategies. To establish a new area of astronomy we need to
go further. In particular, we need to address a
number of immensely challenging modelling problems. Many relevant gravitational-wave
scenarios involve extreme physics, much of which cannot be tested in the laboratory.

Once third generation detectors, like the Einstein Telescope \cite{ET1},
come on-line we should be firmly in the era of gravitational-wave astronomy. These instruments will improve
the broad-band sensitivity by another order of magnitude. This may seem  somewhat remote, given that such detectors are
still at the design stage, but we need to consider their promise now. After all, we need
to argue the case for building these expensive instruments. It is relevant to ask
what we can hope to achieve with the Einstein Telescope, but not (necessarily) with Advanced LIGO.
How much better can we do with (roughly) a one order of magnitude improvement in sensitivity?
Are there situations where this improvement is needed to see the signals in the first place,
or is it just a matter of doing better astrophysics by getting improved statistics and an increased
signal-to-noise to faciliate parameter extraction? There are many interesting issues to consider.

Perhaps in contrast, it is straightforward to argue the case for the space-based detector LISA \cite{lisa}.
Sensitive to low-frequency gravitational waves, LISA is perfectly tuned to typical
astronomical timescales. If the detector works as planned, then detection is guaranteed.
In fact, various known binary systems will be used to verify that the system is operating properly.
The challenges that face the LISA project are different in many ways. Given the number of, in principle,
detectable binaries in the galaxy the data analyst may suffer an embarrassment of riches.
The science may to some extent be confusion limited. However, the fact that LISA is sensitive to
signals from supermassive black holes (either merging or capturing smaller objects) essentially throughout the Universe
makes it an extremely exciting mission. LISA was recently endorsed by the US Decadal Survey, and the hope
is that the detector will fly not too long after 2020.

In 15 years time (say) we should have a network of high-precision instruments searching the skies for
gravitational-wave signals over a range of up to eight decades in frequency. These detectors will provide
us with unprecedented insights into the dark side of the Universe, and allow us to probe much
fundamental physics. Further improvements in data quality is also likely to allow us to measure the level
of the gravitational-wave component in the cosmic microwave background. Ultra-low frequency gravitational
waves may have been detected by pulsar timing arrays.
In parallel, we can expect to have seen breakthroughs in many related areas of physics as
well. The Higgs may have been detected by the LHC, and we may have direct
evidence for supersymmetry. The current controversy concerning possible dark matter signals in the DAMA experiment
should have been resolved, and the technology would have moved on to provide undisputable data. We ought to have a better understanding
of dark energy, e.g. constraints on the ``equation of state'' through the $w$ parameter. These developments will stimulate
theorists as well as experimenters, leading to dramatic improvements in our various models and
ultimately a much better understanding of the Universe in which we live.

\section{Audio \underline{not} video}

Most of the information we have about the Universe has been gleaned from
electromagnetic signals. Ranging from the beautiful high-resolution images
from the Hubble Space Telescope to X-ray timing with  RXTE and
spectra from Chandra, from pulsar timing with radio dishes to the cosmic microwave data from WMAP,
the Sloane Digital Sky Survey and so on. In the last 50 years we have learned that the Universe
is a violent place where stars explode and galaxies collide. There are massive black holes at the centre of most
galaxies, and their evolution (through accretion or mergers) may be closely linked to the actual
formation of structures. The amount of information that we have gathered is truly awesome.
Yet, this Universe is no less mysterious than that of the early 1960s.
As we gather information and improve our understanding of some aspects, there are surprises and
new questions arise. At the present time, detailed questions concern the dynamics of
black holes and their role in evolutionary scenarios, and the state of matter under the extreme
conditions in a neutron star core. The big mysteries concern dark energy and
(obviously) the still uncomfortable marriage between gravity and the quantum.

The ongoing gravitational-wave effort should be considered from this perspective.

From the theory point-of-view, there is a close analogy between
electromagnetic and gravitational waves. However, one must not push it too far.
The two problems are conceptually rather different since electromagnetic radiation
corresponds to oscillations of  electric and magnetic
fields propagating  \underline{in} a given spacetime, while gravitational
waves are oscillations \underline{of} the spacetime itself. An important
upshot of this difference is that one cannot localize the  energy
associated with a gravitational wave
to regions smaller than a wavelength or so.
In order to identify  a gravitational wave
one must identify an oscillating contribution to the spacetime, varying
 on a lengthscale  much smaller
than that of the ``background'' curvature.

Other differences between gravitational and electromagnetic waves
illustrate the promises and challenges of this area:

\noindent
 (i) While electromagnetic waves are radiated when individual particles
are accelerated, gravitational waves are due to non-spherical bulk motion of matter.
In essence, the incoherent electromagnetic radiation generated by many
particles carry information about the thermodynamics of the source.
Gravitational radiation tell us about the large-scale dynamics.

\noindent
(ii) The electromagnetic waves that reach our detectors will have been
scattered many times since their generation. In contrast, gravitational
waves couple weakly
to matter and  arrive at the Earth in pristine condition. They
carry key information about
violent processes that otherwise remain hidden, e.g. associated with the heart of a supernova core collapse
or merging black holes. Of course,
the waves also interact very weakly with our detectors, making detection
a serious challenge.

\noindent
(iii) Standard astronomy is based on deep
imaging of small fields of view, while gravitational-wave
detectors cover virtually the entire sky. A consequence of this is that
their ability to pinpoint a source in the sky is not particularly
good. On the other hand, any source in the sky will in principle
be detectable, not just ones towards which we aim the detector.
This could obviously lead to difficulties if the sources are plentiful.
In fact, this may be a problem for the LISA instrument.

\noindent
(iv) Electromagnetic radiation typically has a wavelength much smaller
than the size of the emitter. Meanwhile, the
wavelength of a gravitational wave is usually comparable to or
larger than the size of the radiating source. Hence,
gravitational waves can not be used for ``imaging''.

The upshot is that gravitational waves carry information about the
most violent phenomena in the Universe; information that is complementary to
(in fact, very difficult to obtain
from) the  electromagnetic data.

\section{Simple estimates}

In  gravitational-wave modelling it is often sufficient to
work at the level of linear perturbations of a given spacetime. That is, we
use
\begin{equation}
g_{\mu \nu} = g_{\mu \nu}^0 + h_{\mu \nu}
\label{perturb}\end{equation}
where $g_{\mu \nu}^0$ is some known background metric and $|h_{\mu \nu}|$ is suitably small.
Massaging the Einstein field equations (by choosing a particularly useful set of ``coordinates'') in vacuum one can show that they reduce to the
 wave equation
\begin{equation}
\Box h_{\mu \nu} = g^{\alpha \beta } \nabla_\alpha \nabla_\beta h_{\mu \nu} =0
\end{equation}
This tells us that changes in the gravitational field propagate as waves, and that these waves travel at the speed of
light. If we consider the effect that these waves of gravity have on matter, we find that they are transverse with two
possible polarizations. Moreover, gravitational waves act like a tidal force which means that they
change all proper distances by the same ratio. Hence, if we consider two
``free masses''  a distance $L$ apart, then the gravitational-wave
induced strain will lead to a change $\Delta L$ in the proper
distance such that
\begin{equation}
h = { \Delta L \over L}
\label{strain}\end{equation}

In order to estimate the expected strength of various gravitational-wave
signal, we can use the well-known flux formula
which relates the luminosity to the strain $h$;
\begin{equation}
{ c^3 \over 16 \pi G} |\dot{h}|^2 = { 1 \over 4 \pi r^2} { dE \over dt}
\end{equation}
where $r$ is the distance to the source. This relation is exact for
the weak waves that bathe the Earth. Let us now
characterize a given event by a timescale
$\tau$ and assume that the signal is essentially monochromatic
(with frequency $f$). Then we can use $dE/dt\approx E/\tau$ and $\dot{h} \approx 2 \pi f h$ and
deduce that
\begin{equation}
h \approx 5\times 10^{-22} \left( { E \over 10^{-3} M_\odot c^2} \right)^{ 1/2}
\left({ \tau \over 1 \mbox{ ms} }\right)^{-1/2}
\left( {f\over 1 \mbox{ kHz}} \right)^{-1} \left( { r\over 15 \mbox{ Mpc}}\right)^{-1}
\label{hraw}\end{equation}
Here we have scaled the  distance to that of a source
in the Virgo cluster. This kind of scaling is necessary to ensure a reasonable event rate
for many astrophysical scenarios. For example, at this distance one would expect to
see many supernovae per year, which means that one can
hope to see a few neutron stars/black holes being born during one year of
observation.

We can proceed to define an ``effective amplitude'' that
reflects the fact that detailed knowledge of the signal can be used to
dig deeper into the detector noise. A typical example is based on the use of matched
filtering, for which the effective amplitude
improves roughly as the square root of the number of observed cycles $N$.
This is a good approximation when $N$ is large, so the estimate will be reliable for
persistent sources but obviously less so for short bursts.
Anyway, using $N \approx f \tau $ we arrive at
\begin{equation}
h_{\rm c}
\approx 5\times 10^{-22} \left( { E \over 10^{-3} M_\odot c^2} \right)^{ 1/2}
\left({ f \over 1 \mbox{ kHz} }\right)^{-1/2}
\left( { r\over 15 \mbox{ Mpc}}\right)^{-1}
\label{heff}\end{equation}
This relation shows us that
the effective gravitational-wave strain, essentially the ``detectability'' of the signal,
depends only on the radiated energy and the characteristic frequency.
This allows us to assess the relevance of a range of proposed sources.

\begin{figure}[h]
\centering
\includegraphics[width=12cm,clip]{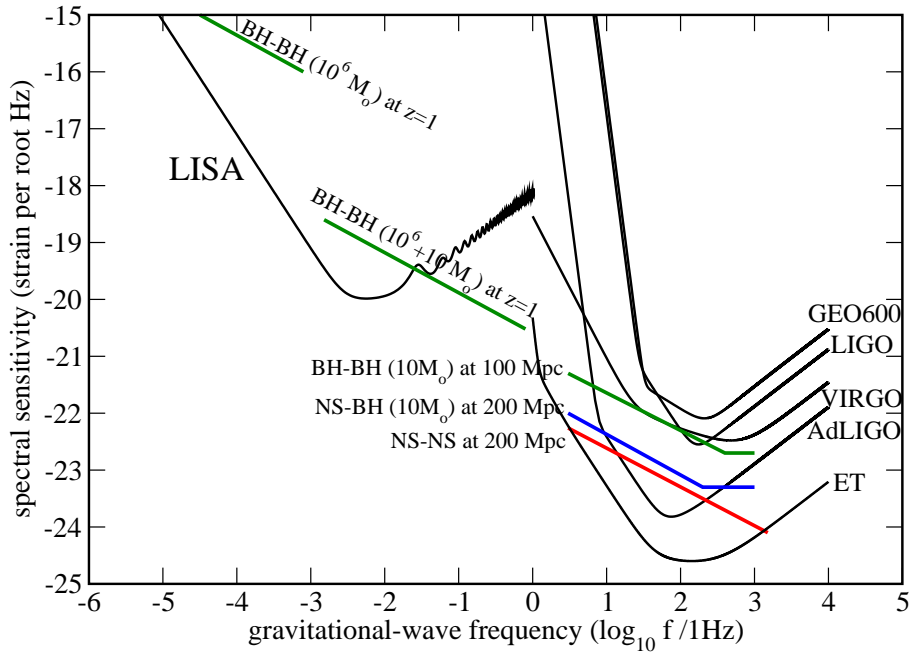}
\caption{Comparing the typical gravitational-wave strain for various binary
systems to the sensitivity of
interferometric detectors. The current generation of ground-based detectors, which are sensitive above 10~Hz,
(GEO600, Virgo and LIGO) is compared to the second generation of detectors, like Advanced LIGO (AdLIGO). Third generation detectors (like the Einstein Telescope, ET) would improve the sensitivity by another order of magnitude across a similar broad frequency range. The figure also shows the space-based detector LISA, which will be a supreme instrument for detecting signals from supermassive black holes. It is worth noting that the dimensionless strain sensitivity of the ground- and space-based detectors is quite similar. To see this, simply multiply the shown curves (showing the strain per root Hz) by the square root of the frequency.}\label{binar}
\end{figure}

To make further progress, we need to develop a better idea of the typical frequencies
associated with different classes of sources. This is fairly straightforward once
we note that the
dynamical frequency of any self-bound system with mass $M$ and radius $R$
can be approximated by
\begin{equation}
f \approx { 1 \over 2 \pi} \sqrt{ {GM \over R^3} }
\label{freq}\end{equation}
Given this,  the natural
frequency of a (non-rotating) black hole should be
\begin{equation}
f_{\rm BH} \approx 10^4 \left( {M_\odot \over M} \right)\mbox{ Hz}
\end{equation}
This immediately shows that medium sized black holes, with masses in
the range $10-100M_\odot$, will be prime sources for ground-based
interferometers, since the "sweet spot" of these detectors tends to be located around 100~Hz, see Figure~\ref{binar}.
Meanwhile,  supermassive black holes
with masses $\sim10^6M_\odot$  expected
to exist in the cores of many galaxies should radiate in
the LISA band. We also see that neutron stars, with a
canonical mass of $1.4 M_\odot$ compressed inside a radius of
10~km or so, would be expected to radiate at
\begin{equation}
f_{\rm NS} \approx 2 \mbox{ kHz}
\end{equation}
In other words, they require detectors that are sensitive at high frequencies.

Finally, we need to consider binary systems. In that case we
can (roughly) take $R$ to represent the separation between the two objects.
If we assume that the final plunge sets in at something like $R\approx 6M$ (in geometric units $G=c=1$)
we find that a binary system with  two $10 M_\odot$ black holes
will radiate at frequencies below 1.5~kHz or so. In other words,
such systems should be interesting sources for ground-based
detectors. Again, the argument tells us that supermassive black-hole binaries --- resulting from
galaxy mergers --- radiate in the LISA frequency band.
The frequency range of the space-based interferometer
(down to $10^{-4}$~Hz) is also a good match to the timescale of known
astronomical systems (hours). There are many classes of
galactic binary systems that radiate gravitationally in the
LISA band and which should lead to detectable signals.
Typical such systems are i) binary white dwarfs, ii)
binaries comprising an accreting white dwarf and a Helium donor star,
iii) low-mass X-ray binaries.
In fact, there may be more than $10^8$ galactic
binaries in the LISA frequency range.

These back-of-the-envelope estimates provide a sketch of the scope
of gravitational-wave astronomy.
Of course, they only tell us if more detailed modelling is worthwhile.
Unfortunately, the next step tends to very difficult, either involving poorly understood physics
(as in the case of neutron stars) or complex nonlinear dynamics (as for black-hole collisions), or indeed both (as
in the case of neutron star mergers and core collapse).
These requirements have led to the development of numerical relativity as a high-powered tool
for astrophysical simulations. At the same time a range of issues bridging
nuclear physics, particle physics and field theory, low-temperature physics and hydrodynamics
relevant for neutron stars are being investigated. Fundamental physics associated with the
early Universe or the dark matter/energy models in modern cosmology is also under vigorous scrutiny.

\section{Promises and challenges}

We will now consider a set of sample problems that provide interesting modelling challenges
for the future. The chosen problems (obviously) do not provide a complete list in any sense.
Rather, they have been selected to illustrate particular aspects and give an idea of the big picture.

\subsection{Compact binary inspiral (and merger)}

It is natural to start by considering  inspiralling binaries. While we are still waiting
for the first undisputed detection of gravitational waves, compact binaries provide the strongest
support for Einstein's theory. The indirect evidence from double neutron star systems is clear. After more than 30 years of detailed monitoring, we know that the orbital decay of
the famous binary pulsar 1913+16 decays at a rate that agrees with the predictions of general relativity to within a
fraction of a percent. Hence, gravitational waves are a reality --- not a ghost. Similar binary systems,
much later in their evolutionary history, are expected to provide the first direct detection as well.

In contrast to the case in Newtonian gravity the two-body problem remains
unsolved in General Relativity. Given the lack of suitable exact
solutions to the Einstein field equations significant effort has gone
into developing various approximations and numerical approaches to the problem.

Within the post-Newtonian approximation
the leading order radiation effects are described by the quadrupole formula, according to which
the gravitational-wave strain follows from the second time derivative of
the source's quadrupole moment
\begin{equation}
Q_{jk} = \int \rho x_j x_k dV \rightarrow Q \sim { MR^2}
\end{equation}
in such a way that
\begin{equation}
h_{jk} = { 2G \over rc^4} { d^2 Q_{jk} \over dt^2} \rightarrow h \sim { M^2 \over rR}
\label{quadro}\end{equation}
Here we have (for simplicity) assumed that we are considering an equal mass binary
with separation $R$.

As the system emits gravitational waves, it loses energy and the orbit shrinks.
This leads to an evolutionary timescale;
\begin{equation}
t_{\rm chirp} \sim { R^4 \over M^3}
\end{equation}
Assuming that this timescale is shorter than the observation time (which obviously means that we
are considering the final stages before merger), we
can estimate,  using (\ref{freq}), the effective amplitude of the binary signal as
\begin{equation}
h_c \approx \sqrt{ f t_{\rm chirp}}\  h \sim { M \over r} \left( {R\over M}\right)^{1/4}
\end{equation}
This shows that even though the raw signal gets stronger (as the frequency chirps up towards its cut-off value at plunge and merger) its detectability decreases as the orbit shrinks.
Hence, we need the detectors to be sensitive at low
frequencies where the
binary system spends a lot of time.

The estimates also explain why
binary systems are particularly attractive to observers; the amplitude of the signal
is ``calibrated'' by the two masses.

Ground-based detectors should be able to track a neutron star binary as it evolves from
a few Hz to coalescence, radiating around $10^4$ cycles in the process.  This means that
one can enhance the detectability by roughly a factor of 100 if one can track the
signal through the entire evolution (without losing a single cycle of the wave-train). This motivates the development of
high order post-Newtonian approximations to the waveforms \cite{pn} (especially the phase),
as well as detailed signal analysis algorithms \cite{data}.

Any binary system
which is observable from the ground will coalesce within one year.
Statistics based on the pulsar population then tells us that these events would happen
less than once every $10^5$~yrs in our Galaxy. Hence, we
need to detect  events from a volume of space containing at least
$10^6$ galaxies in order to see a few mergers per year. Translating this into distance using our understanding of the
mass distribution in the Universe, we see that a detector must
be sensitive enough to see coalescing binaries
beyond a few hundred Mpc in order for the event rate to be reasonable.

Observers often discuss the detectability of binary systems in terms of the
 ``horizon'' distance $d_h$ at which a given signal would be observable with a particular detector.
Let us assume that detection requires a signal-to-noise ratio of 8, and focus on equal mass neutron star
binaries (each star has mass $1.4 M_\odot$). For such systems, $d_h$ would be 30~Mpc for the LIGO S5 data.
In this volume of space one would expect one event every 25-400 yrs. Advanced LIGO should improve this to
$d_h=300$~Mpc, and could
see several to hundreds of events per year, while the Einstein Telescope
may reach $d_h=3$~Gpc and should potentially observe many thousands of events.

These estimates tell us why, first of all, it would have been surprising to find a
binary signal in the S5 data. Given even the most optimistic rate estimate from
population synthesis models, these events would be rare in the observable volume of space.
The situation changes considerably with Advanced LIGO; one would expect
neutron star binaries to be seen once the detectors reach this level of sensitivity. However, it is also clear
that if the most pessimistic rate estimates are correct, then we will not be able to gather a statistically significant sample of signals. Most likely, we will need detectors like the Einstein Telescope to study populations.

Third generation detectors will also be required if we want to study the final stages of inspiral, including the
actual merger. This is a very interesting phase of the evolution given that
the merger will lead to the formation of a hot compact remnant with violent dynamics. Since the stars are magnetized,
the merger may also trigger a
long gamma-ray burst. Most of this dynamics radiates at relatively high frequencies. Tidal disruption
occurs above 600~Hz or so and the oscillations of the remnant could lead to a signal at
several kHz. However, the merger signal should be rich in information. In particular, the
ringdown should tell us directly whether a massive neutron star or a black hole was formed.
Roughly, if the inspiral
phase is observable with Advanced LIGO then the Einstein Telescope should be able to detect the merger. In other words, the
development of third generation detectors is essential if we want to study these events.
In parallel, we need to improve our models of the merger phase. This requires nonlinear
simulations in full general relativity, accounting for the complex physics associated with a
high-density/extremely hot remnant. Realistic models would also have to include magnetic fields,
describe neutrino emission and cooling etcetera. Definite progress is being made towards this goal, but
severe challenges remain.

The situation is quite different for black-hole binaries. The last few years have seen a breakthrough
in numerical relativity, to the point where the problem of inspiralling and merging black holes
is considered ``solved''. This progress is of immense importance as it provides experimenters with reliable
templates that can be used to develop optimal data analysis strategies.

Black-hole systems are of obvious interest for gravitational-wave physics.
Importantly, a black-hole binary should be more massive than one comprising neutron stars, and hence
it will lead to a stronger gravitational-wave signal. We expect to see much more distant
black-hole binaries, ultimately out to cosmological distances. Of course, we do not at this point have any evidence that such binaries actually exist in the Universe.
Stellar evolution theory tells us that they ought to be out there, but we do not know this for a fact.
Gravitational-wave observations should change this.

More detailed calculations show that in the case of unequal masses the
leading order signal depends only on the so-called ``chirp-mass'', the
combination $\mu^{3/5} M_T^{2/5}$
where $\mu$ is the reduced mass and $M_T$ the total mass. If one observes the
shrinking time of the orbit as well as the gravitational-wave amplitude then
one can infer the chirp mass and  the
distance to the source. This means that
coalescing binaries are ``standard candles'' which
may be used to infer the Hubble constant
and other cosmological parameters \cite{cosmo}.
By extracting higher order post-Newtonian terms one can hope to infer
the individual masses, the spins and maybe also put constraints on the
graviton mass.

In the case of LISA, one would expect to be able to observe mergers of supermassive
black holes with very high signal-to-noise ratios (several 1000$^{\rm s}$).
This means that one may be able to see such events
no matter where in the Universe they occur. The information gained from such observations
will shed light on the evolution of these gigantic black holes, via a sequence of mergers or accretion, and
improve our understanding of the development of large-scale structures in the Universe.

Another key problem for LISA concerns the capture of smaller
bodies by large black holes \cite{emri}. The space-based detector should be able to detect many such events.
Their detailed signature provides
information about the nature of the spacetime  in the vicinity of the black hole,
and could therefore be used to map the geometry, test black-hole uniqueness theorems etcetera.
To model these systems is, however, far from trivial.
In particular
since the orbits may be highly eccentric. The main challenge concerns the
calculation of the effects of radiation reaction on the smaller body.
In addition to accounting for the gravitational self-force and the radiation reaction, one must develop
a computationally efficient scheme for modelling actual orbits. This is not easy, but at least we
know what the key issues are.

\subsection{Supernovae}

Our expectations are not always brought out by more detailed modelling. Sometimes the devil is in the
detail and our intuition falters. For example,
one might expect apparently powerful events like supernova explosions and
the ensuing gravitational
collapse to lead to very strong gravitational-wave
signals. However, the outcome depends entirely on the asymmetry
of the collapse process. This is clear from \eqref{quadro}. The difference between the initial and the final state does not matter. It is the route that the system takes, \underline{how} it evolves, that determines the strength of the gravitational-wave signal. Unfortunately, current numerical simulations
 suggest that the level of radiation from core collapse events is rather low. Typical
results suggest that an energy equivalent to $\sim 10^{-7} M_\odot c^2$
(or less!) will be radiated \cite{ott}.

Combining the anticipated energy release with the typical dynamical timescale for a collapsing compact core,
around a millisecond (frequency $\sim$1~kHz), we learn from Eq.~(\ref{heff})
that the gravitational-wave amplitude may be of the order of
 $h_c  \sim  10^{-22}$
for a source in the Virgo cluster.
This estimate (which accords reasonably well with full numerical
simulations) suggests that these
sources are unlikely to be detectable beyond the local group
of galaxies. This would make observable events very rare. It is
expected that three to four supernovae will go off every century
in a typical galaxy. This means that we would be very lucky to see
one in our Galaxy given only a decade or so of observation.
However, one should keep in mind that even a single unique
event would provide great insights into supernova physics.
In fact, the gravitational waves carry unique
information away from a collapsing system, and the detailed signature may allow us to distinguish between
different proposed explosion mechanisms \cite{etpaper}. While the optical signal emerges
hours, and the neutrino burst several seconds, after the collapse,
the gravitational waves are generated during the collapse itself.
This means that the gravitational-wave signal carries a clean
signature of the collapse dynamics. This information may be impossible to
extract in any other way.

\subsection{Spinning neutron stars}

When the dust settles from the supernova event we are left with a compact remnant,
either a neutron star or a black hole of a few solar masses. As we have already seen,
both classes of objects are relevant to the gravitational-wave physicist.
Black holes involve extreme spacetime curvature, and will allow us to probe
strong field aspects of Einstein's theory. Meanwhile,
neutron stars are cosmic laboratories of exciting physics that cannot
the tested by terrestrial experiments.
With a mass of more than that of the Sun compressed inside a radius of about
10 kilometers, their density reaches beyond nuclear saturation.
We already have a wealth of data from radio, X-ray and gamma-ray observations, providing
evidence of an incredibly rich phenomenology. We know that neutron stars
appear in many different guises,  from radio pulsars
and magnetars to accreting millisecond pulsars, radio transients and intermittent
pulsars. Our models for these systems remain rather basic, despite 40 years of attempts to
understand the pulsar emission mechanism, glitches, accreting systems etcetera.

Importantly, neutron stars can radiate gravitationally through a number of different mechanisms \cite{etpaper}.
Relevant scenarios include the binary inspiral and merger that we have already discussed,
rotating stars with deformed elastic crusts, various modes of oscillation and a range of
associated instabilities.
Modelling these different scenarios is not easy since the physics
of neutron stars is far from well known.
To make progress we must
combine supranuclear physics (the elusive equation of state) with magnetohydrodynamics,
the crust elasticity, a description of
superfluids/superconductors and potentially also exotic phases of matter involving a
deconfined quark-gluon plasma or hyperons. Moreover, in order to be quantitatively accurate, the models
have to account for relativistic gravity.

Much effort has been invested in understanding the rich spectrum of oscillations of
``realistic'' neutron star models. This is natural since such oscillations may be excited to a relevant
level at different stages in a neutron star's life. Gravitational waves from a pulsating neutron star could provide an
excellent probe of the star's properties, and may allow us to infer
the mass and radius with good precision.
This would obviously help constrain the supranuclear equation of state,
which is currently known only up to factors of two or so.
The most promising scenarios involve unstable oscillations \cite{modes}.
As an unstable pulsation mode grows it may reach a sufficiently
large amplitude that the emerging gravitational waves can be detected.
In the last decade the inertial r-modes have been under particular scrutiny, following the
realization that they are particularly prone to a gravitational-wave induced instability.
This is an exciting and active area of current research but we will nevertheless not
go into details here. The problem is covered elsewhere in this volume \cite{kostas}.
Instead, we will focus on rotating deformed neutron stars.

As soon as a newly born neutron star cools below roughly $10^{10}$~K
(within a few minutes)
its outer layers begin to crystallize; freezing to form
the neutron star crust. The crust is not very rigid --- in fact, it is rather
jelly-like --- but it can still sustain shear stresses.
Asymmetries in the crust, expected to arise due to its evolutionary history, will slowly leak
rotational energy away from  a spinning neutron star. Such sources would be the
gravitational-wave analogue of
radio pulsars, radiating at twice the spin frequency.
On the one hand, rotating neutron stars will emit low amplitude
waves, but on the other hand, they radiate continuously for a long time. Moreover, we have many potential target
sources with known frequency and position. This means that observers can carry out
a targeted search for known radio and X-ray pulsars.

It is straightforward to estimate the signal strength for this kind of sources.
Expressing the asymmetry in terms of an
ellipticity $\epsilon$, we find that
\begin{equation}
h\approx 8\times 10^{-28} \left( { \epsilon \over 10^{-6}} \right)
\left( {f \over 100 \mbox{ Hz} }\right)^2
\left( { 10 \mbox{ kpc} \over r} \right)
\label{bar}\end{equation}
where we have used the fact that the gravitational-wave frequency $f$
is twice the  rotation frequency (and assumed a canonical $M=1.4M_\odot$
and $R=10$~km neutron star). The source distance has been scaled for objects in our galaxy,
since this is all we can hope to detect anyway.
This signal is still far too weak to be detected directly, but
the effective amplitude increases (roughly) as the square-root of the
number of detected cycles. Accounting for this and assuming an
observation time of one year, we need
\begin{equation}
\epsilon > 2.5 \times 10^{-6} \left( {100 \mbox{ Hz} \over f} \right)^{5/2}
\left( { r \over 10 \mbox{ kpc}} \right) \left( { h_c \over 10^{-22}} \right)
\label{epseff}\end{equation}
Combined with the expected detector sensitivities, i.e. some idea of the achievable $h_c$,
this estimate allows us to assess whether
a given deformation is likely to be detectable.

Of course, this is a purely academic exercise unless we have some idea of the
level of asymmetry one would expect a neutron star to have. This is
a complicated problem, where the answer depends not only on the properties of the
star, but also on its evolutionary history. So far, modelling
has mainly focused on establishing what the largest possible neutron star ``mountain'' would be.
The most detailed models suggests that
$\epsilon < 2\times 10^{-5} \left( {u_\mathrm{break}/ 0.1} \right)$.
Recent molecular dynamics simulations suggest that the breaking strain is $u_\mathrm{break} \approx 0.1$, much
larger than originally anticipated. In comparison to
terrestrial materials, which have $u_\mathrm{break} \approx 10^{-4} - 10^{-2}$,  the
neutron crust may be super-strong!

In this context, observations of targeted radio pulsars are already providing interesting results.
Based on
1 month of S3/4 data LIGO set the strong constraint $\epsilon <7\times10^{-7}$ for J2124-3358 \cite{def2}. This tells us that this, relatively fast spinning, pulsar is far from maximally deformed.
An observational milestone was later reached when S5 data was used to beat the Crab pulsar ``spin-down limit'' by a factor
of 4 or so. This limit has since been improved to the gravitational-wave energy being less than a few percent of the
total energy release \cite{def1}. In other words, gravitational-wave emission does not dominate the spin-down of these systems.
Although this was already ``known'' from observed pulsar braking indices, it is clear that the
gravitational-wave searches are beginning to produce astrophysically relevant results.

It is quite easy to estimate how these results are likely to improve in the future since the
effective amplitude of a periodic signal increases as the square root of the observation time.
In the case of J2124-3358 one would expect analysis of the S5 data (with a factor 2 improved sensitivity, and a full year of data) to improve the constraint to $\epsilon<10^{-7}$. Advanced LIGO, with an order
of magnitude better sensitivity, but still  a one year integration, should reach $\epsilon<10^{-8}$, and
the Einstein Telescope may push the limit as far as $\epsilon<10^{-9}$. At this point, the deformation of the star would be constrained to
the micron level. One would probably expect a signal to be detected before this level is reached, but we do
not know this for sure. The main issue concerns the generation of the deformations.
Why should  the neutron star be deformed in the first place?
This is an urgent problem that needs to be addressed
by theorists.

As far as evolutionary scenarios are concerned, accreting neutron
stars in low-mass X-ray binaries have attracted the most attention. This is natural for a number of reasons.
First of all, the currently observed spin-distribution in these systems seems consistent with the presence
of a mechanism that halts the spin-up due to accretion well before the neutron star reaches the break-up limit \cite{nature}.
Gravitational-wave emission could provide a balancing torque. The required
deformation is certainly smaller than the allowed upper limit. Using the standard accretion torque
we find that we need
\begin{equation}
\epsilon \approx 4.5\times10^{-8} \left( { dM/dt \over 10^{-9} M_\odot/\mbox{yr}} \right)^{1/2}
\left( {300 \mbox{ Hz} \over \nu_s} \right)^{5/2}
\end{equation}
Moreover, in an accreting system it would be quite natural
for asymmetries to develop. However, accreting systems are very messy, and we do not
understand the accretion torque very well. Hence, reality could be quite different from
our rough estimate. Another issue concerns the need to integrate the signal for a long time to build
up the signal-to-noise ratio. Given the general behaviour of accreting neutron stars it is not clear that we
will be able to track such systems (and integrate coherently) for long enough \cite{anna}. Basically, we need to
invest more effort in trying to understand these systems. Having said that, we have every reason to expect
progress on the key issues. After all, the evolution of accretion X-ray binaries is also high on the agenda
for our colleagues in other parts of astrophysics.

\subsection{Fundamental physics/cosmology}

Arguably, the most fundamental observation we can hope to make with future gravitational-wave detectors
would be a cosmological background from the Big Bang. Asymmetries in the very early Universe would be amplified by the expansion, resulting in a broad gravitational-wave spectrum in the present Universe. The slope of this
spectrum and possible peaks provide information about masses of particles, energies of relevant phase-transitions, and perhaps even the sizes of key dimensions. Detection of such stochastic signals is rather different
from the problems that we have considered so far. In particular, it requires cross-correlation of the output from several detectors. This provides a new set of challenges.

Current results
constrain the energy density of the stochastic gravitational-wave background (normalized to the critical energy density of the Universe) around 100~Hz to be less than $7\times10^{-6}$ \cite{stoch}. At this level, some extreme theoretical suggestions are confronted but the standard scenarios remains very safe.

Unfortunately, the gravitational waves associated with the
standard inflationary scenario are out of reach for the planned ground-based instruments and LISA.
However, LISA's frequency band ($\sim$mHz) represents gravitational waves
that had the horizon size at the electroweak phase-transition. If this transition were first order, then there could be a detectable stochastic background. Another relevant possibility involves cosmic strings, which emit
gravitational waves with a characteristic signature \cite{strings}. These waves may be detectable even if they do not make a significant contribution to the mass budget of the Universe. However, the best search window for such signals, essentially free of ``local'' gravitational-wave sources, is around 0.1---1 Hz.  This basically means that we need a LISA follow-up mission in space. Obviously, any such venture is not going to take place soon.

The planned detectors may nevertheless make significant contributions to our understanding of fundamental
physics. Sometimes opportunities arise serendipitously, as in the case of the possibility to use GEO600 to test ideas based on the holographic principle \cite{holo}. In other cases, fundamental physics is closely linked with planned observations.
As an example of this, consider the fact that we expect to be able to infer distances to coalescing binary systems, providing a distance scale of the Universe in a precise calibration-free measurement. This idea will be pursued with the next generation of ground-based detectors in the first place, but their reach is obviously not as impressive as that of LISA and the Einstein Telescope. In fact,
LISA has fantastic sensitivity to massive black hole mergers at $z=1$ and would be able to detect
$10^4 M_\odot$ systems out to $z=20$. Let us suppose that these merger events have an observable electromagnetic counterpart. This would provide us with redshift information, and as a consequence we may use LISA data
to constrain the dark energy equation of state parameter $w$ (and perhaps probe $dw/dt$).
In other words, LISA may also be relevant as a dark energy mission \cite{cosmo}. This is an interesting idea, providing yet another
example of the interdisciplinary nature of gravitational-wave physics.

\section{Multi-messenger astronomy}

Gravitational-wave astronomy promises to shed light on the ``dark side'' of the Universe.
Because of their strong gravity, black holes and neutron stars are ideal
sources and we hope to probe the extreme physics associated with them.
The potential for this is clear, in particular given third generation detectors (like the Einstein Telescope)
and the space interferometer LISA. However, in order to detect the signals and extract as much information as possible, we need to improve our
theoretical models considerably.

The binary black-hole problem may have been ``solved'' by the recent progress in numerical relativity,
but for neutron stars binaries many issues remain.
We need to work out when finite
size effects begin to affect the evolution. We need to consider tidal resonances and compressibility
in detail and ask to what extent they affect the late stages of inspiral. For hot young remnants,
resulting from binary mergers or core collapse, we need to refine our nonlinear simulations. The
simulations must use ``realistic'' equations of state, and consider composition,
heat/neutrino cooling and magnetic fields with as few ``cheats'' as possible. Similar issues obviously arise
for supernova core-collapse simulations, which set the current standard for including realistic physics. In parallel, we need to
improve our current understanding of neutron star oscillations and instabilities. This effort should aim
at accounting for as much of the interior physics as possible. Finally, we need a clearer
phenomenological understanding of pulsar glitches, accreting neutron stars, magnetar flares etcetera.
These are ambitious targets, but there is no reason why we should not make good progress in the next few years.
Eventually, observations (gravitational and electromagnetic) will  help us understand
many aspects of extreme physics that seem mysterious to us today.

The next decade will see significant
improvements in the various observational
channels, and we should expect great progress
in our understanding of the Universe.
Gravitational-wave physics, with its
promise to provide information that is
truly complementary to electromagnetic
observations, has an important part to play
in this enterprise. In fact, the precision
modelling and sophisticated data analysis
tools that are essential for the gravitational-wave
experiments, should be valuable also
for future X-ray and radio astronomy.
Evidence for this is clear from the emergence
of numerical relativity simulations as a
reliable tool for studying violent astrophysical
phenomena. These developments may
have been motivated by gravitational-wave
physics, but they are finding applications
in a wider context. One would expect
these kinds of synergies to continue to
develop, particularly as many gravitational-wave
sources may have electromagnetic
counterparts. Perhaps the most exciting
results will come from a combination
of gravitational-wave, electromagnetic and
neutrino channels --- true multimessenger astronomy.

\section*{Acknowledgements (and apologies)}

As is no doubt clear, this is not a detailed review of this subject. Based on an overview talk at the Astroparticle Physics School in Erice 2010, this article draws on a number of sources and builds on the work of many colleagues.
Shamelessly, I decided not to even attempt to cite most of the original sources. To make amends for this, I would like to collectively thank everyone that have helped me develop my current understanding of this exciting science.
Everything that I get wrong is, of course, my own fault...


\begin{thebibliography}{99}
\bibitem{holo} An interesting story that is  best followed up by a web search for ``holographic noise'' and ``GEO600''.
\end{thebibliography}
\end{document}